\documentclass[10pt, conference, compsocconf]{IEEEtran}
\usepackage[utf8]{inputenc} 
\usepackage[T1]{fontenc}
\usepackage{url}
\usepackage{ifthen}
\usepackage{cite}
\usepackage[cmex10]{amsmath}
\usepackage{amsthm}

\usepackage{graphicx}
\usepackage{amsfonts}
\usepackage{amssymb}
\usepackage{mathrsfs}
\usepackage{subfigure}
\usepackage{mdwmath}
\usepackage{mdwtab}
\usepackage{dsfont}
\usepackage[ruled,vlined,linesnumbered]{algorithm2e} 

\newtheoremstyle{quest}
  {5pt}
  {5pt}
  {}
  {0pt}
  {\bfseries}
  {.}
  { }
  {\thmname{#1}\thmnumber{ #2}\thmnote{ (#3)}}
\theoremstyle{quest}

\newtheorem{theorem}{Theorem}
\newtheorem{lemma}{Lemma}

\newtheorem{defi}{Definition}

\newenvironment{iarray}{\begin{IEEEeqnarray}{rCl}}{\end{IEEEeqnarray}\ignorespacesafterend}

\begin{document}

\title{Optimizing Information Freshness in Broadcast Network with Unreliable Links and Random Arrivals: An Approximate Index Policy
}

\author{\IEEEauthorblockN{Jingzhou Sun$^1$, Zhiyuan Jiang$^2$, Sheng Zhou$^1$, Zhisheng Niu$^1$}
    \IEEEauthorblockA{$^1$Beijing National Research Center for Information Science and Technology,\\Department of Electronic Engineering, Tsinghua University, Beijing 100084, China.\\
        $^2$Shanghai Institute for Advanced Communication and Data Science, Shanghai University, Shanghai 200444, China.\\
        sunjz18@mails.tsinghua.edu.cn, zhiyjiang@foxmail.com, \{sheng.zhou, niuzhs\}@tsinghua.edu.cn,\\}}

    \maketitle

\begin{abstract}
With the rapid growth of real-time Internet of Things (IoT) applications, the need for fresh information has surged. Age of Information (AoI) is a tailor-made metric to characterize the information freshness perceived by the devices. In this paper, we investigate the problem of scheduling updates to minimize AoI in broadcast network. In this case, a central controller, e.g. a base station, collects status updates from different sources and schedules them to corresponding clients. Different from previous work, we consider both stochastic status updates and unreliable links. The problem is first modeled as an infinite horizon average constrained cost Markov decision problem (CMDP). With Lagrangian relaxation, an approximation of Whittle's index is derived and a scheduling policy is designed based on the approximate index. The results in previous work can be view as degenerate cases of the approximation index policy either with reliable links or periodic arrival constraints. Simulation results demonstrate the near-optimal performance of the proposed policy.
\end{abstract}


\section{Introduction}

Recent years have witnessed a runaway rise in the number of mobile devices and applications which boosted the need for real-time information. These applications include safety applications in intelligent transportation system\cite{1} and state synchronization for robust electric power network \cite{2}. 
The ensuing problem is how to measure the freshness of information. Towards this end, \emph{Age of information} (AoI)  was proposed recently in \cite{3} to quantify the information observation delay at the destination, which is formally defined as the time elapsed since the generation time of the last received update. It is a tailor-made performance metric for status update systems, wherein the freshness of status information is of critical importance in decision making at the destination.


Extensive efforts have been devoted to minimizing AoI under various scenarios in the literature. With one server considered, queuing theory has been applied in \cite{4,5} to analyzing and optimizing AoI by adjusting service rate and arrival rate under various queuing model. Age Minimization problem in multi-hop network has been investigated in \cite{6,7}. The authors in \cite{8,9,10} study how to minimize peak AoI and average AoI in general interference network and provide the relation of peak AoI and average AoI. 


In this paper, we consider the problem of link scheduling to minimize AoI in a broadcast network where packets carrying time sensitive status information arrive at the scheduling center randomly and the transmission channels are unreliable. There have been a number of noticeable advances accomplished recently in this problem based on the Whittle's index approach \cite{11} and optimal packet management of keeping the freshest packet only. In \cite{kadota18}, Kadota \emph{et al.} derived closed-form index expression considering unreliable transmissions but deterministic packet arrivals (every $T$ time slot). Hsu \cite{hsu18} generalized it to stochastic packet arrivals but assuming no-buffer terminals and reliable channel. Our previous work \cite{jiang18_isit,jiang18_itc} derived the Whittle's index for stochastic packet arrivals with arbitrary buffer size but the transmissions are assumed reliable. To the best of our knowledge, no explicit index expressions are available for scenarios with double randomness, i.e., random arrivals and unreliable links. The coexistence of random arrivals and unreliable links renders the problem particularly challenging because channel condition and AoI must be considered together. Our contribution is to develop an approximate index policy for this double-random broadcast network. Besides, we find that the results in \cite{kadota18,jiang18_itc} coincide with special cases of our result. Simulations are carried out to demonstrate the performance of the scheduling policy based on the approximate index.

\section{System Model and Problem Formulation}
\label{sec_sm}
\begin{figure}[!t]
\centering
\includegraphics[width=0.4\textwidth]{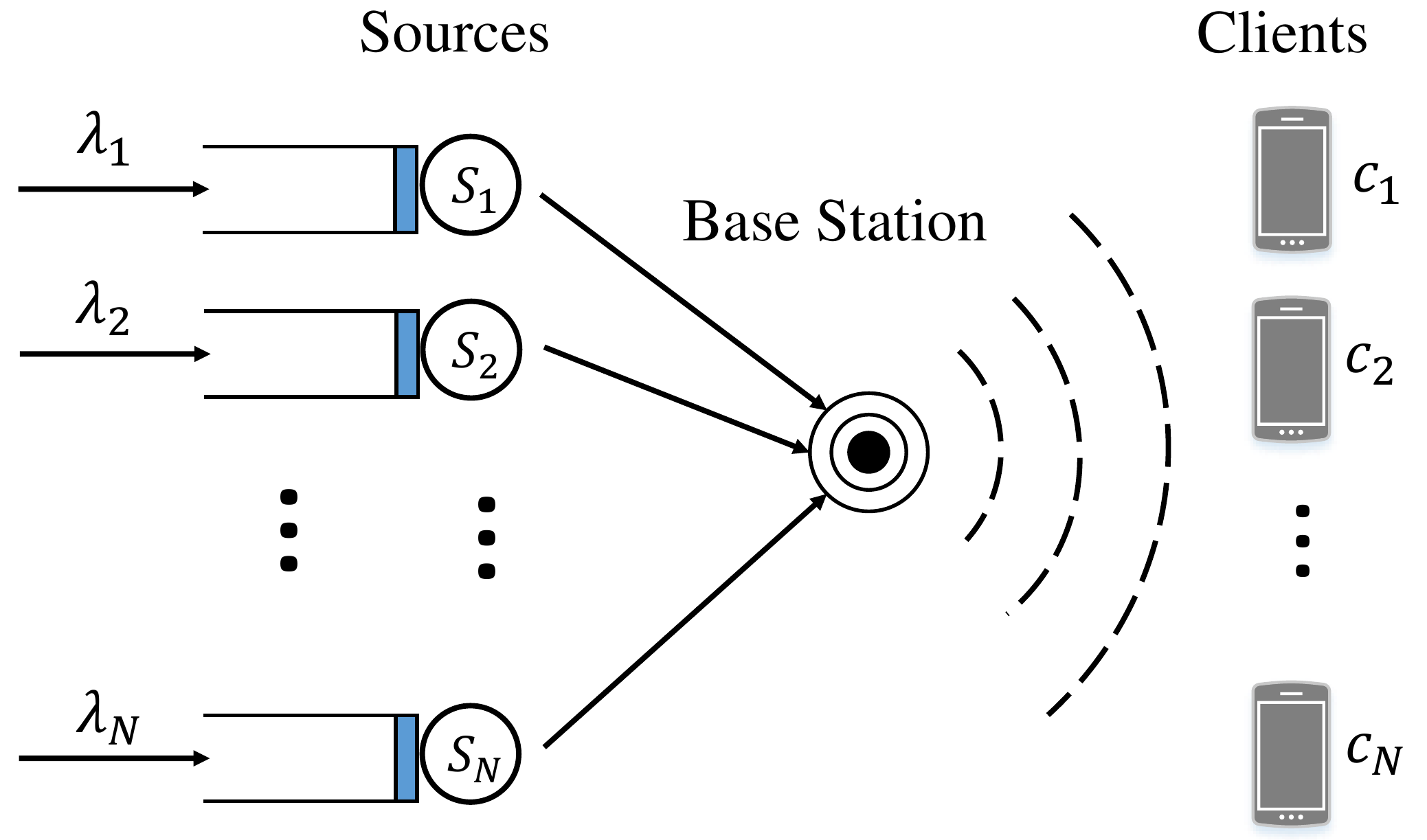}
\caption{Broadcast Network with a base station broadcasting information from one source each time. The status updates from sources are stochastic and the links between base station and clients are unreliable.}
\label{Fig_System_Model}
\end{figure}
We consider a wireless broadcast network depicted in Fig.\ref{Fig_System_Model}, where one base station (BS) collects status information from sources and transmits information to clients. Let time be slotted and the BS can only broadcast information from one source in each time slot. We assume that the client $u_i$ requests information from the corresponding source $s_i$ which transmits the information via BS. This peer-to-peer relation can be viewed as a generalization of other cases. Considering an extreme case, for example, where one client receives information from $N$ sources, we can unpack this client into $N$ virtual clients each collecting information from one source, and these two are equivalent. 

The number of packets from source $s_i$ arriving at the BS in time slot $t$ is denoted as $\Lambda_i\left( t\right)\in \{0,1\}$, where $\Lambda_i\left( t\right) = 1$ means there is one packet arriving at the BS. Assuming the distribution of $\Lambda_i\left( t\right)$ is time-invariant, 
then let $\lambda_i = \mathbb{E}\{\Lambda_i\left( t\right)\}$. Since our main concern is how to schedule down link transmission, we assume the generation process cannot be adjusted and the packets can be transmitted immediately to the BS via wired network once generated. At the beginning of each time slot, the BS can transmit at most one packet and finish it within the time slot. Due to the uncertainty of wireless channel, the transmission may fail. Let $\Gamma_i\left( t\right) \in \{0,1\}$ be the channel state of the link between BS and the client $u_i$.  If $\Gamma_i\left( t\right) =1$, transmission through this link will succeed. Otherwise, the transmission must fail if $\Gamma_i\left( t\right) =0$. Supposing the channel state is a Bernoulli random variable, we only know the statistical information $p_i = \mathbb{E}\{\Gamma_i\left( t\right)\}$. The scheduling decision at time slot $t$ is defined as a $N$-dimension vector $\mathbf{U}(t) = (u_1(t),\cdots,u_N(t))$. Each entry $u_i(t)$ is an indicator, and $u_i(t) = 1$ means the BS decides to transmit a packet to client $c_i$ at time slot $t$. A scheduling policy $\pi$ is composed of the scheduling decision at each time slot.

The AoI at client $c_i$ at time $t$ is denoted by $A_i\left( t\right) $. The value of $A_i\left( t\right) $ is the time elapsed since the generation time of the last received packet by client $c_i$. To minimize time average AoI, it is obvious that the one-buffer packet management is optimal, where the BS only keeps the newest packet for each client and all the other undelivered packets for this client are discarded. Let $a_i\left( t\right) $ be the queuing delay of the packet buffered for client $c_i$ at time $t$. Then $a_i\left( t\right)$ is equal to the time elapsed since the generation time of the buffered packet. We assume that the scheduling happens at the beginning of each time slot while new arrival happens at the end of it. The dynamics of the AoI can be written as
\begin{equation}
    A_{i}\left( t+1\right) =
    \begin{cases}
    a_{i}\left( t\right) +1,&\text{if $u_i(t)\Gamma_i\left( t\right) =1$},\\
    A_{i}\left( t\right) +1,&\text{otherwise,}
    \end{cases}
\end{equation}
and
\begin{equation}
    a_{i}\left( t+1\right) =
    \begin{cases}
    1,&\text{if $\Lambda_i\left( t\right) =1$},\\
    a_{i}\left( t\right) +1,&\text{otherwise.}
    \end{cases}
\end{equation}
 Here, $a_i\left( t\right)  = 1$ for newly arriving packet because the increment of AoI happens right after packet arrival.

The average AoI under policy $\pi$ is defined as
\begin{equation}
\label{AoI}
    \Delta_{\pi} \triangleq \limsup_{T\to\infty}\frac{1}{T N}\mathbb{E}_{\pi}\left[\sum_{t=1}^T \sum_{i=1}^N A_{i}\left( t\right) \right].
\end{equation}
Our problem is how to design a policy $\pi$ with low complexity to minimize $\Delta_{\pi}$. This problem can be modeled as an infinite time horizon average cost constrained Markov decision process (CMDP) with $\{a_1\left( t\right) ,A_1\left( t\right) ,\cdots,a_N\left( t\right) ,A_N\left( t\right) \}$ as the state space. The constraint is that the BS transmits one packet in each time slot. Generally, the optimal scheduling policy can be obtained by value iteration or policy iteration for finite state CMDP. However, the state space of the propose CMDP is countably infinite. To cope with it, we resort to the Whittle's index method under which the CMDP is decoupled into sub-problems with smaller state space. A low complexity scheduling policy is designed based on Whittle's index.

\section{Scheduling Algorithm Design}
In this section, we first introduce the basic concept of Whittle's index method \cite{11}. Then we apply this method to our problem to derive an approximate Whittle's index and design the scheduling policy.
\subsection{Whittle's Index}

Considering the CMDP mentioned above, the main challenge is that the states of different clients are intertwined. An intuitive idea is to decouple this problem into many sub-problem through Lagrangian relaxation, where the decoupled model only involves one client. In each time slot, the BS chooses whether to transmit to this client or not, denoted by $u\in \{1(active), 0(passive)\}$. The optimal policy is to minimize the total cost of scheduling this client. Let $u^*(a,d) = 1$ be the optimal action under state $(a,d)$, and $\mathcal{A} = \{(a,d): u^*(a,d) = 1\}$ and $\mathcal{P} = \{(a,d): u^*(a,d) = 0\}$.


Whittle's index measures how attractive it is to transmit under given state by \emph{subsidy for passivity}. In the decoupled model, a Lagrangian multiplier is added to the one step cost, which can be interpreted as a subsidy. To put it clear, a constant subsidy $W$ is obtained whenever the action is passive. The introduction of subsidy changes the active set and passive set. Let $\mathcal{P}(W)$ be the paasive set of states under subsidy $W$. If $\mathcal{P}(W)$ monotonically increases from $\varnothing$ to the whole state space as $W$ goes from $0$ to $\infty$, then the CMDP is call $\emph{indexable}$. Under indexable condition, the minimum subsidy $W$ needed to move state $(a,d)$ from active set to passive set measures how attractive this state is. This minimum $W(a,d)$ is called the \emph{Whittle's index} of state $(a,d)$. Scheduling policy based on Whittle's index is to transmit to client with the largest Whittle's index. Whereas Whittle's index policy is suboptimal for minuscule cases, it is believed that this scheduling policy is near optimal as the total number of clients goes to infinity in most cases if the problem is indexable\cite{17}.

Unfortunately, the indexability of the CMDP is often difficult to establish, though it may seem intuitive. Computing Whittle's index can also be complex, often rely on numerical approximations. A closed-form Whittle's index may not exist even though the problem is indexable.

\subsection{Decoupled Model}
The decoupled model can be naturally formulated as a MDP. We first introduce the state, action, one step cost, transition, and objective of this MDP. For convenience, the client index is omitted since only one client is considered. The MDP's state is $s\left(t\right) = \left(a\left(t\right),d\left(t\right)\right)$, where $d\left(t\right) = A\left(t\right)-a\left(t\right)$. The action is denoted by $u\left(t\right)\in\{0,1\}$, and $u\left(t\right) = 1$ indicates the BS tries to update the client, otherwise the BS idles. We consider poster-action cost which depends on the states after taking action. Besides, subsidy $W$ is deducted from the cost when the BS idles. 
the transition probability is given by
\begin{iarray}
&& \Pr\{\left(a,d\right) \to \left(a+1,d\right) \} = \left(1-\lambda\right)\left(1-p\right); \nonumber\\
&& \Pr\{\left(a,d\right) \to \left(1,d + a\right) \} = \lambda\left(1-p\right); \nonumber\\
&& \Pr\{\left(a,d\right) \to \left(1,a\right) \} = \lambda p; \nonumber\\
&& \Pr\{\left(a,d\right) \to \left(a+1,0\right) \} = \left(1-\lambda\right)p.
\end{iarray}
And when $u\left(t\right) = 0$, the probability becomes
\begin{iarray}
&& \Pr\{\left(a,d\right) \to \left(a+1,d\right) \} = 1-\lambda; \nonumber\\
&& \Pr\{\left(a,d\right) \to \left(1,d + a\right) \} = \lambda.
\end{iarray}
The objective function under policy $\pi$ is
\begin{iarray}
\label{origin}
J_{\pi} = \limsup_{T\to\infty}\frac{1}{T}\sum^{T}_{t=1}\mathbb{E}_{\pi}\left[a\left(t\right)+d\left(t\right)-\left(1-u\left(t\right)\right)W\right]
\end{iarray}

By verifying the conditions in \cite{16}, it can be proved that a stationary and deterministic policy is cost-optimal. Thus, we omit the time index and present the Bellman equation
\begin{iarray}
\label{Bellman}
h\left(a,d\right)+J = \min \{\mu_0\left(a,d\right),\mu_1\left(a,d\right)\}
\end{iarray}
where
\begin{iarray}
\label{mu_content}
&&\mu_0\left(a,d\right) = \lambda h\left(1,a+d\right)+\left(1-\lambda\right)h\left(a+1,d\right) +a+d-W; \nonumber\\
&&\mu_1\left(a,d\right) = \left(1-p\right)\left[\lambda h\left(1,a+d\right)+\left(1-\lambda\right)h\left(a+1,d\right)\right]  \nonumber\\
&&+p\left[\lambda h\left(1,a\right)+\left(1-\lambda\right)h\left(a+1,0\right)\right]+a+\left(1-p\right)d. 
\end{iarray}
$J$ is the optimal average cost and $h\left(a,d\right)$ is the cost-to-go function. In \cite{jiang18_itc}, we proved that when $p=1$, 
\begin{equation}
\label{mon}
    h\left(a,0\right) \le h\left(a,1\right) \le...\le h\left(a,d\right) \le...
\end{equation}
This property can be explained intuitively that $h\left(a,d\right)$ gets larger as the AoI of the client increases. It can be proved that this property still holds when $p<1$ by considering the corresponding discount problem and details are omitted here. Based on the property, the cost optimal policy can be specified to be threshold-type.
\begin{defi}
If policy $\pi$ satisfies that the action of state $\left(a,d\right)$ is to idle when $d\le D_a$ and to update when $d\ge D_a$, then this policy is \emph{threshold-type}.
\end{defi}
\begin{theorem}
\label{th1}
There exists a threshold-type policy that is cost optimal for the decoupled model. 
\end{theorem}
\begin{IEEEproof}
Since 
\begin{iarray}
\mu_1\left(a,d\right)-\mu_0\left(a,d\right) =&& p\left(1-\lambda\right)\left[h\left(a+1,0\right)-h\left(a+1,d\right)\right] \nonumber \\
&& +p\lambda\left[h\left(1,a\right)-h\left(1,a+d\right)\right]\nonumber\\
&& +W-pd,
\end{iarray}
is a non-increasing function of $d$ according to the property mentioned above. Suppose the optimal action under state $\left(a,d\right)$ is to idle, which means $\mu_1\left(a,d\right)-\mu_0\left(a,d\right)\ge 0$, then 
\begin{align*}
\mu_1\left(a,d-1\right)-\mu_0\left(a,d-1\right)&&\ge  \mu_1\left(a,d\right)-\mu_0\left(a,d\right)\ge 0. 
\end{align*}
Thus the optimal action of state $(a,d-1)$ is also to idle and this establishes the theorem.
\end{IEEEproof}

Considering the threshold $\{D_a\}$, we proved that these thresholds are monotonic when $p = 1$ \cite{jiang18_itc},
\begin{equation}
\label{th}
    D_1 \le D_2 \le...\le D_a \le...
\end{equation}
The monotonic property of $\{D_a\}$ is based on the intuition that the BS is more inclined to wait for a new packet arrival as the queuing delay $a$ increases instead of updating. When $p<1$, this property can also be established by considering the discounted MDP. If $D_a$ is an integer, we have
\begin{iarray}
\label{core}
\mu_1\left(a,D_a\right)=\mu_0\left(a,D_a\right),
\end{iarray}
which means the two actions are equivalent because $D_a$ is the threshold.

To derive the Whittle's index, we should compute $\{D_a\}$ at first based on \eqref{core}. Towards this end, it is necessary to compute the cost-to-go function $h(a,d)$. When $p=1$ and $d\ge D_a$, we find that $h(a,d)$ is invariant w.r.t. $d$ based on \eqref{Bellman} and \eqref{mu_content}. Unfortunately, when $p<1$, all these states are deeply intertwined which renders it particularly difficult to derive a close-form Whittle's index. Thus, we resort to its approximation.

\begin{theorem}
\label{th2}
For the decoupled model, given subsidy $W$, the threshold $\{D_a\}$ of the optimal policy satisfy,
\begin{iarray}
D_a
    \begin{cases}
    \le\frac{a-1}{\Delta}\left(\frac{W}{p D_1}+\frac{D_1+1-a}{2}-\Delta\right)+D_1,\quad \text{$a \le D_1$},\\
    \le \frac{W}{p\Delta},\quad \text{$D_1\le a \le \frac{W}{p\Delta}$},\\
    = \frac{W}{p\Delta},\quad \text{$a \ge \frac{W}{p\Delta}$},
    \end{cases}
\end{iarray}
where 
\begin{iarray}
\label{eq24}
D_1 &&\le \sqrt{\frac{2W}{p}+\left(\Delta-\frac{1}{2}\right)^2} -\left(\Delta-\frac{1}{2}\right),\nonumber\\
\Delta &&= \frac{1}{\lambda}+\frac{1-p}{p}.\nonumber
\end{iarray}
The proof is presented in Appendix \ref{app1}.
\end{theorem}

Let $\{D'_a\}$ be the upper bound of the actual thresholds $\{D_a\}$ shown in \eqref{eq24}. 
Recall that the Whittle's index of state $(a,D_a)$ is $W$. If we use the surrogate $\{D'_a\}$ and let the approximate index value of $(a,D'_a)$ be $W$, then the approximate index of state $(a,d)$ is,
\begin{iarray}
\label{eq25}
W\left(a,d\right)=
    \begin{cases}
    \frac{p}{2}x^2+p\left(\Delta-\frac{1}{2}\right)x,\quad \text{if $\frac{d\Delta}{a}\ge\frac{a-1}{2}+\Delta$,}\\
    pd\Delta,\quad \text{otherwise.}
    \end{cases}
\end{iarray}
where
\begin{iarray}
\label{eq25}
x\triangleq \frac{d\Delta +\frac{a\left(a-1\right)}{2}}{a-1+\Delta}.\nonumber
\end{iarray}
When the problem is indexable, the approximate Whittle's index is smaller than the actual Whittle's index. From this expression, we notice $W(a,d)$ can be approximated as a linear function of $p$ when $p$ approaches $1$. This indicates that clients with smaller $p$ are more likely to experience stale information 
It should also be remarked that the approximate  index is consistent with previous work. Letting $p=1$ or $\lambda =1$, the approximate index degenerates to the index proposed in \cite{jiang18_itc} or \cite{kadota18}, which means that the approximate index is of high quality in heavy traffic or high reliable channel cases.

To guarantee the near optimal performance of Whittle's index theoretically, it remains to prove that the problem is indexable. When $p=1$, we proved it in \cite{jiang18_itc}. Although it is intuitively right when $p<1$, we find it is really hard to establish indexability. However, simulation results show that the approximate index policy still boasts near optimal performance.

\section{Simulation}
\begin{figure}[!t]
\centering
\includegraphics[width=0.47\textwidth]{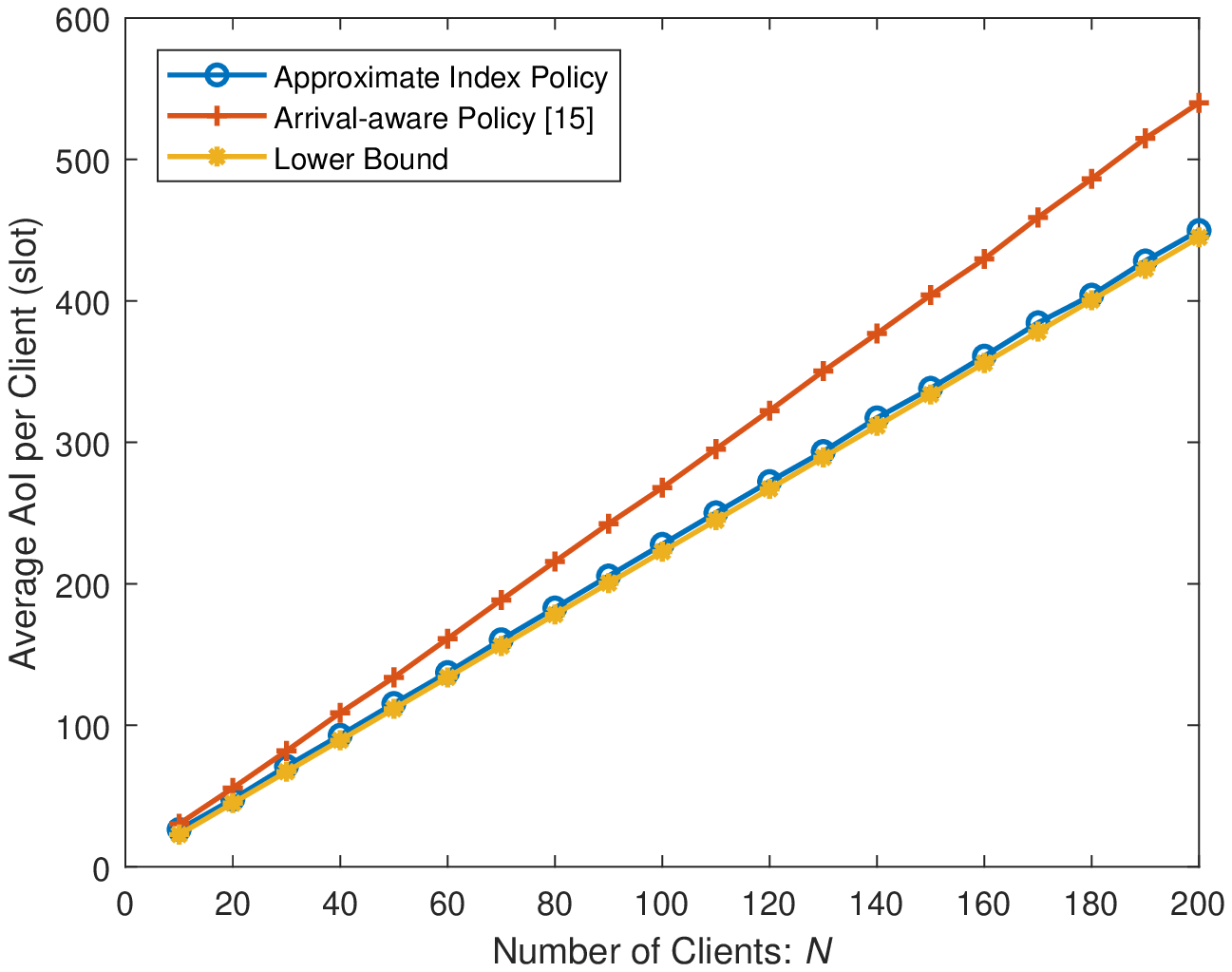}
\caption{Network with $\lambda = \frac{10}{N+10}$. Half of the clients have $p_1 = 0.9$, the rest have $p_2 = 0.1$, }
\label{simu_fig_1}
\end{figure}

In this section, we evaluate the performance of the approximate index. We compare it with policy in \cite{jiang18_itc}. The policy in \cite{jiang18_itc} only considers the randomness of the arrival process and we call it Arrival-Aware Policy. 
Assuming the arriving probability $\lambda$ is 1 and the variance of the inter-delivery time is zero,  we derive a lower bound $L_B$ of the average AoI as in \cite{kadota18},
\begin{iarray}
L_B = \frac{1}{2N}\left(\sum_{i=1}^{N}\frac{1}{\sqrt{p_i}}\right)^2+\frac{1}{2}
\end{iarray}

\begin{figure}[!t]
\centering
\includegraphics[width=0.47\textwidth]{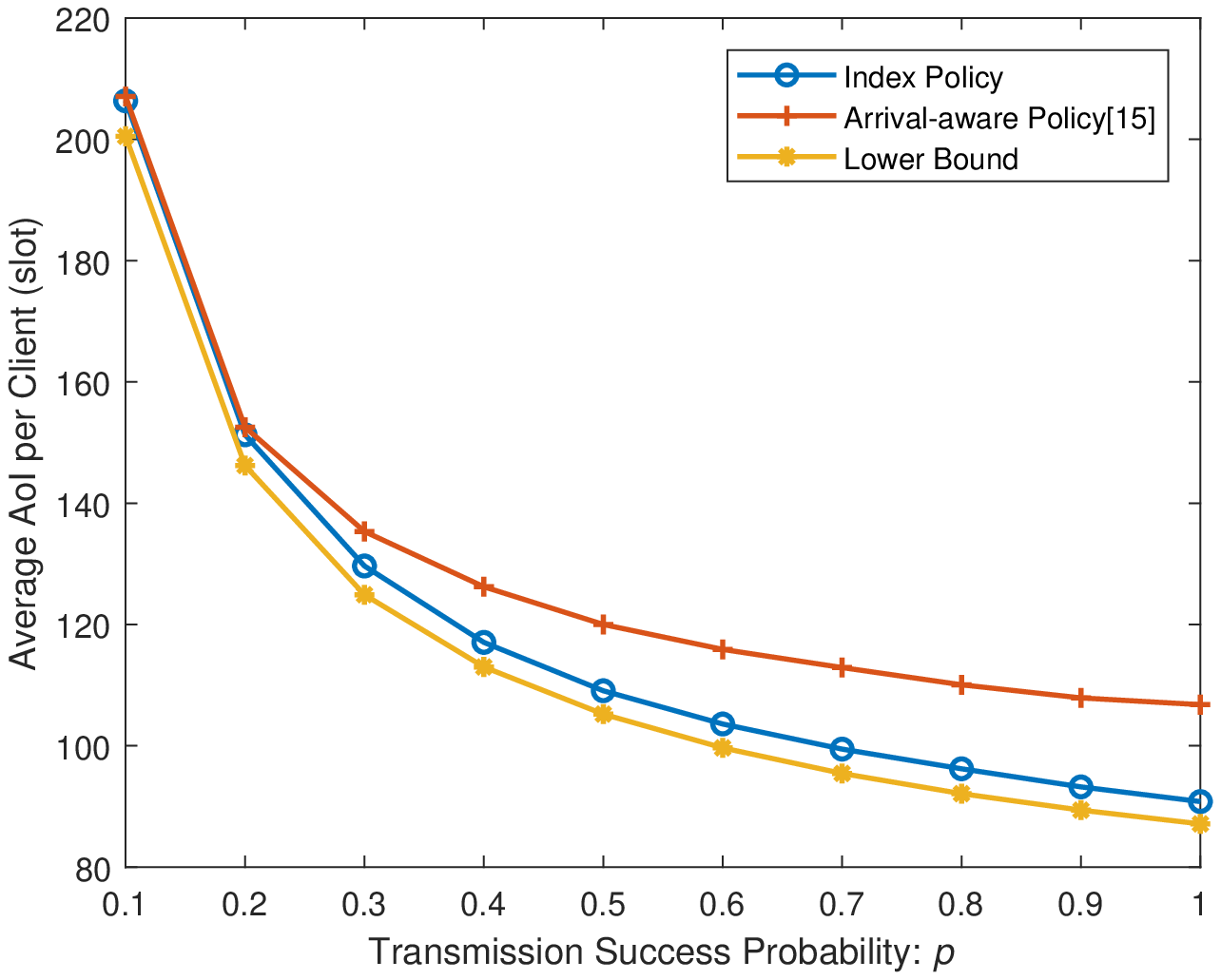}
\caption{The number of clients is $N=40$. Arrival probability $\lambda = 0.2$. Half of the clients have fixed $p_1 = 0.1$. The rest have $p\in\{0.1,0.2,\cdots, 1\}$ }
\label{simu_fig_2}
\end{figure}

Fig.\ref{simu_fig_1} shows the performance of scheduling polices in large network with multiple clients, e.g. the number of of clients $N\in\{10,20,\cdots,200\}$. The arrival probability is $\frac{10}{N+10}$. The channels of half of the clients are of high quality with $p=0.9$ and the rest channels are poor with $p =0.1$. The simulation running time slot is $6N\times 10^4$. As we can see, the performance of our approximate index policy is comparable to the lower bound in this setting. The gap between arrival-aware policy and approximate index policy grows as the number of clients increases and this represents the benefit of taking the randomness of wireless channel into consideration.

Fig.\ref{simu_fig_2} evaluates the performance under different transmission success probability. Here, $N=40$ and the simulation running time is $3\times 10^6$ slot. The arrival probability is $0.2$. The transmission success probability for half of the clients is fixed and its value is $0.1$. For the rest client, the probability is $p_i\in\{0.1,0.2,\cdots,1\}$. When $p_i$ is close to $0.1$, the two scheduling policies have the same performance, because the probability is the same for all clients. As $p_i$ increases, two lines diverges. The gap between the lower bound and approximate index policy mainly comes from the fact that the lower bound is not tight.

\section{Conclusions}
\label{sec_con}
This paper considers a time-slotted network with one BS sending time-sensitive status information to multiple clients. We developed a low complexity scheduling policy to minimize the average Age of Information. Compared to the previous work \cite{jiang18_itc,kadota18,hsu18}, this work is the first to consider both the randomness of arrival process and unreliable transmission. We find that the scheduling priority can be approximated as a linear function of the transmission success probability in high reliable channel case. Through simulations, we observe that the performance of our scheduling policy is close to the lower bound of average AoI under various arrival probability and heterogeneous transmission success probability. Future work includes performance analysis of the policy and considering time-varying channels and other stochastic arrival process. Distribute scheduling algorithm to minimize AoI is also an interesting topic. 

\section*{Acknowledgement}
This work is sponsored in part by the Nature Science Foundation of China (No. 61871254, No. 91638204, No. 61571265, No. 61861136003, No. 61621091), National Key R\&D Program of China 2018YFB0105005, and Hitachi Ltd.

\appendices
\section{Proof for Theorem \ref{th2}}
\label{app1}
To obtain the structure the thresholds, we first prove the following lemmas. Without loss of generality, we assume the domain of the bias $h\left(a,d\right)$ is extended to real number field, such that $\mu_1\left(a,D_a\right)=\mu_0\left(a,D_a\right)$ always holds even when $D_a$ is not an integer.
\begin{lemma}
\label{lm1}
There exists $D^*>0$ such that $\lim_{a\to\infty} D_a = D^*$
\end{lemma}
\begin{IEEEproof}
When $pd\ge W$,
\begin{iarray}
    && \mu_0\left(a,d\right) - \mu_1\left(a,d\right) \nonumber\\
    &=& pd-W + p\left(1-\lambda\right)\left(h\left(a+1,d\right)-h\left(a+1,0\right)\right) \nonumber\\
    && + p\lambda\left(h\left(1,d+a\right)-h\left(1,a\right)\right) \ge 0.
\end{iarray}
which indicates that the optimal action is to update when $d\ge \frac{W}{p}$. Since $D_a$ is monotonic, there must exists $D^*>0$ such that $\lim_{a\to\infty} D_a = D^*$.
\end{IEEEproof}

\begin{lemma}
\label{lm2}
For any $a_1,a_2\ge1$, $0 \le d_1\le D_{a_1}$, $0 \le d_2\le D_{a_2}$, and $a_1+d_1=a_2+d_2$, we have 
\begin{equation}
    h\left(a_1,d_1\right)=h\left(a_2,d_2\right). 
\end{equation}
\end{lemma}

\begin{IEEEproof}
For simplicity, let $K = a_1 + d_1$.
Since $d_1\le D_{a_1}$, the optimal action of state $\left(a_1,d_1\right)$ is to idle. Hence,
\begin{iarray}
\label{eq1}
    h\left(a_1,d_1\right) + J &=& \mu_0\left(a_1,d_1\right) \nonumber\\
    &=& \lambda h\left(1,K\right)+\left(1-\lambda\right) h\left(a_1+1,d_1\right)\nonumber \\
    && +K.
\end{iarray}
According to \eqref{th}, the optimal action of state $\left(a_1+n,d_1\right)$ is still to idle, when $n\ge 0$. Therefore,
\begin{iarray}
\label{eq2}
    h\left(a_1+n,d_1\right) + J =&& K+n + \lambda h\left(1,K+n\right)\nonumber \\
    && +\left(1-\lambda\right) h\left(a_1+n+1,d_1\right).
\end{iarray}
Replacing the second term in \eqref{eq1} with \eqref{eq2} recursively, we obtain
\begin{iarray}
\label{eq3}
    h\left(a_1,d_1\right) &=& \sum_{i=0}^{n}\left(1-\lambda\right)^{i}\left(K+i-J\right)\nonumber\\
    && +\sum_{i=0}^{n}\lambda\left(1-\lambda\right)^{i} h\left(1,K+i\right)\nonumber \\
    && + \left(1-\lambda\right)^{n+1} h\left(a_1+n+1,d_1\right).
\end{iarray}
Let $n\to\infty$,
\begin{iarray}
\label{eq4}
    h\left(a_1,d_1\right) &=& \sum_{i=0}^{\infty}\left(1-\lambda\right)^{i}\left(K+i-J\right)\nonumber \\
    && +\sum_{i=0}^{\infty}\lambda\left(1-\lambda\right)^{i} h\left(1,K+i\right).
\end{iarray}
It can be found that the value of $h\left(a_1,d_1\right)$ only depends on $K$. Thus, when $a_1+d_1=a_2+d_2$, $h\left(a_1,d_1\right)=h\left(a_2,d_2\right)$.
\end{IEEEproof}

\begin{lemma}
\label{lm3}
When $a\ge 1$, 
\begin{equation}
\label{eq5}
    h\left(a,D_a\right)-h\left(a,0\right)=\frac{W}{p}.
\end{equation}
\end{lemma}

\begin{IEEEproof}
Since
\begin{iarray}
\label{eq6-1}
    \mu_0\left(a,D_a\right)=\mu_1\left(a,D_a\right)
\end{iarray}
According to \eqref{Bellman}, we obtain
\begin{iarray}
\label{eq6}
    h\left(a,D_a\right)-h\left(a,0\right) = \mu_1\left(a,D_a\right)-\mu_0\left(a,0\right)
\end{iarray}
Expanding $\mu_1$ and $\mu_0$ according to \eqref{mu_content} and rearranging these terms in \eqref{eq6-1} and \eqref{eq6}, we have
\begin{iarray}
\label{eq7}
    h\left(a,D_a\right)-h\left(a,0\right) &=& \frac{W}{p}.
\end{iarray}
\end{IEEEproof}

\begin{lemma}
\label{lm4}
When $d\ge D^*$, 
\begin{equation}
\label{eq8}
    h\left(a,d+1\right)-h\left(a,d\right)=\frac{1-p}{p}, \quad\hfill\square
\end{equation}
\end{lemma}

\begin{IEEEproof}
Let $G\left(a,d\right)=h\left(a,d+1\right)-h\left(a,d\right)$. Based on \eqref{th} and Lemma 1, when $d\ge D^*$,
\begin{iarray}
\label{eq9}
    G\left(a,d\right) &=& 1-p+\left(1-p\right)\lambda G\left(1,a+d\right)\nonumber\\
    && +\left(1-p\right)\left(1-\lambda\right) G\left(a+1,d\right)
\end{iarray}
Expanding the last term recursively, 
\begin{iarray}
\label{eq12}
    G\left(a,d\right) &=& \lambda\left(1-p\right)\sum_{i=0}^{\infty}\left(\left(1-p\right)\left(1-\lambda\right)\right)^i G\left(1,a+d+i\right)\nonumber \\
    && + \frac{1-p}{p+\lambda -p\lambda}
\end{iarray}
Therefore,
\begin{iarray}
\label{eq15}
    G\left(1,d\right)-\left(1-p\right)\left(1-\lambda\right)G\left(1,d+1\right) &=& \lambda\left(1-p\right)G\left(1,d+1\right)\nonumber \\
    && +1-p
\end{iarray}
and this yields
\begin{iarray}
\label{eq16}
    G\left(1,d\right)=1-p+\left(1-p\right)G\left(1,d+1\right)
\end{iarray}
Solving this equation, we get $G\left(1,d\right) = \frac{1-p}{p}$. Combining with \eqref{eq12}, we obtain that $G\left(a,d\right) = \frac{1-p}{p}$.
\end{IEEEproof}

\begin{lemma}
\label{lm5}
\begin{iarray}
\label{eq18}
    h\left(a+1,0\right)-h\left(a,0\right) = \frac{1}{\lambda}+\frac{1-p}{p},\quad \forall a\ge D^*. 
\end{iarray}
\end{lemma}

\begin{IEEEproof}
When $a\ge D^*$,
\begin{iarray}
\label{eq19}
    h\left(a+1,0\right)-h\left(a,0\right) &=& 1+\lambda\left(h\left(1,a+1\right)-h\left(1,a\right)\right)\nonumber\\
    && \left(1-\lambda\right)\left(h\left(a+2,0\right)-h\left(a+1,0\right)\right)\nonumber\\
    &\overset{a}{=}& \left(1-\lambda\right)\left(h\left(a+2,0\right)-h\left(a+1,0\right)\right)\nonumber \\
    && +1+\frac{\left(1-p\right)\lambda}{p}
\end{iarray}
The equation $(a)$ is based on Lemma 4.
Solving \eqref{eq19} recursively yields that
$h\left(a+1,0\right)-h\left(a,0\right) = \frac{1}{\lambda}+\frac{1-p}{p}$
\end{IEEEproof}

\begin{lemma}
\label{lm6}
The limit $D^*=\frac{\lambda W}{\lambda +p-p\lambda}$, and $D_a=D^*$ when $a\ge D^*$. 
\end{lemma}

\begin{IEEEproof}
When $a\ge D^*$,
\begin{iarray}
\label{eq17}
    h\left(a+D_a,0\right)-h\left(a,0\right) &=& \left(\frac{1}{\lambda}+\frac{1-p}{p}\right)D_a.
\end{iarray}
From Lemma 2, $h\left(a+D_a,0\right) = h\left(a,D_a\right)$. Combining it with Lemma 3 yields
\begin{iarray}
\label{eq17}
    \left(\frac{1}{\lambda}+\frac{1-p}{p}\right)D_a = \frac{W}{p} \quad \forall a\ge D^*
\end{iarray}
This means that $D_a$ is constant when $a\ge D^*$. Since the thresholds $\{D_a\}$ converge to $D^*$, we obtain that $D^* = \frac{\lambda W}{\lambda +p-p\lambda}$.
\end{IEEEproof}



When $1\le a\le D_1$, we have
\begin{iarray}
\label{eq20-1}
    h\left(a,0\right) &=& a-J+\lambda h\left(1,a\right)+\left(1-\lambda\right)h\left(a+1,0\right)
\end{iarray}
Because $a\le D_1$, $h\left(1,a\right)=h\left(a+1,0\right)$. Therefore, 
\begin{iarray}
\label{eq20}
    h\left(a,0\right) = a-J+h\left(a+1,0\right)
\end{iarray}
With $h\left(1,0\right) = 0$, we obtain 
\begin{iarray}
\label{eq21-1}
    h\left(a,0\right) = \left(a-1\right)J-\frac{a\left(a-1\right)}{2}~\forall a\le D_1+1
\end{iarray}
From \eqref{eq19}, the difference satisfies
\begin{iarray}
\label{eq21}
    h\left(a+1,0\right)-h\left(a,0\right) \ge \frac{1}{\lambda}+\frac{1-p}{p} \quad\forall D_1\le a \le D^*
\end{iarray}
Let $a=D_1$ in \eqref{eq20} and \eqref{eq21}, we get
\begin{iarray}
\label{eq22}
J-D_1\ge \frac{1}{\lambda}+\frac{1-p}{p}.
\end{iarray}
According to Lemma \ref{lm2} and \ref{lm3}, 
\begin{iarray}
h\left(D_1+1,0\right) = h\left(1,D_1\right) = h\left(1,0\right)+\frac{W}{p} = \frac{W}{p},
\end{iarray}
and $h\left(D_1+1,0\right) = D_1 J-\frac{D_1\left(D_1+1\right)}{2}$. Thus,
\begin{iarray}
\label{eq23}
D_1 J-\frac{D_1\left(D_1+1\right)}{2}=\frac{W}{p}.
\end{iarray}
Solving \eqref{eq22} and \eqref{eq23} gives
\begin{iarray}
\label{eq24}
D_1\le \sqrt{\frac{2W}{p}+\left(\Delta-\frac{1}{2}\right)^2} -\left(\Delta-\frac{1}{2}\right),
\end{iarray}
where $\Delta = \frac{1}{\lambda}+\frac{1-p}{p}$.

When $a\le D_1$, 
\begin{iarray}
\label{eq25}
h\left(a+D_a,0\right) &&\ge h\left(D_1+1,0\right)+\left(a+D_a-D_1-1\right)\Delta; \nonumber \\
h\left(a+D_a,0\right) &&= h\left(a,D_a\right) = h\left(a,0\right)+\frac{W}{p}.
\end{iarray}
And this yields,
\begin{iarray}
\label{eq26}
D_a \le \frac{a-1}{\Delta}\left(\frac{W}{p D_a}+\frac{D_1+1-a}{2}-\Delta\right)+D_1.
\end{iarray}
When $D_1\le a\le D^*$, 
\begin{iarray}
\label{eq27}
h\left(a+D_a,0\right) &&\ge h\left(a,0\right)+D_a\Delta; \nonumber \\
h\left(a+D_a,0\right) &&= h\left(a,D_a\right) = h\left(a,0\right)+\frac{W}{p}.
\end{iarray}
Thus, when $D_1\le a\le D^*$,
\begin{iarray}
\label{eq28}
D_a \le D^*.
\end{iarray}

\bibliography{Reference.bib}
\bibliographystyle{IEEEtran}
\end{document}